\documentclass[12pt,a4paper]{article}

\usepackage[utf8]{inputenc}
\usepackage{amssymb}
\usepackage{multicol}
\usepackage[margin=2cm]{geometry}
\usepackage{setspace}
\usepackage{graphicx}
\usepackage{url}
\usepackage{color}
\usepackage{hyperref}
\graphicspath{ {images/} }
\usepackage{tikz}
\usepackage{amsmath}
\usepackage{physics}
\usepackage{ulem}
\usepackage{xcolor}
\usepackage{centernot}
\usepackage{amsthm}

\usepackage{fancyhdr}

\usepackage{centernot}
\usepackage{braket}

\begin{document}

\title{\textbf{An Alleged Tension between Non-classical Logics and Applied Classical Mathematics}}

\doublespacing

\author{Sebastian Horvat\thanks{Faculty of Physics, University of Vienna} \and Iulian D. Toader\thanks{Institute Vienna Circle, University of Vienna}}

\date{ }

\maketitle

\begin{abstract} Timothy Williamson has recently argued that the applicability of classical mathematics in the natural and social sciences raises a problem for the endorsement, in non-mathematical domains, of a wide range of non-classical logics. We first reconstruct his argument and present its restriction to the case of quantum logic (QL). Then we show that there is no problematic tension between the applicability of classical mathematical models to quantum phenomena and the endorsement of QL in the reasoning about the latter. Once we identify the premise in Williamson's argument that turns out to be false when restricted to QL, we argue that the same premise fails for a wider variety of non-classical logics. In the end, we use our discussion to draw some general lessons concerning the relationship between applied logic and applied mathematics.

\bigskip


\bigskip

\textbf{Word count}: 8985

\end{abstract}

\newpage 

\section{Introduction}
The recent philosophical literature on logic has seen a significant rise in views that reject some of the epistemological and metaphysical tenets that have traditionally been attributed to logic, such as its alleged aprioricity, analyticity and necessity (see e.g. Martin \& Hjortland, 2022, and references therein). It has also been maintained that choosing a logical theory proceeds along similar abductivist lines to how theories are chosen in the empirical sciences (Priest, 2016), and more particularly, that developments in the empirical sciences themselves may (or even should) inform our logical beliefs (Williamson, 2017).\footnote{For a criticism of abductivism about logic, see Hlobil, 2021.} A prominent defender of such a view, Timothy Williamson, has recently put his logical-abductivist convictions in action by advancing an argument - henceforth referred to as \textit{Williamson's argument} or, for short, \textbf{WA} - to the effect that the way mathematics is used in the natural and social sciences provides abductive support for classical logic over some of its non-classical competitors. 

The class of logics specifically targeted by \textbf{WA} includes all those logics motivated by reasons \textit{external} to mathematics, as exemplified by quantum and many-valued logics, which indeed have their origins in empirical considerations. Logics that reject some classically-valid argument or law due to issues \textit{internal} to mathematics -- like the intuitionist's rejection of the law of excluded middle, motivated by constructivist views about mathematics -- are not within the targeted class. To be sure, it is not a logic's actual origin that is of concern here, but rather its intended domain of application, i.e. the piece of discourse whose logical structure it purports to adequately capture. Nevertheless, together with Williamson, we will switch interchangeably between a logic's origin and its intended domain of application, as the two typically coincide. It is also worth noting that \textbf{WA} has nothing to say against so called \textit{pure logics}, that is, unapplied mathematical objects that carry `logic' in their name only because they resemble the mathematical structures of classical logic. 

The aim of the present paper is to show that \textbf{WA} is unsound, to explain why we think this is so, and to expand on what we think there is to learn from this explanation concerning the general relationship between applied logic and applied mathematics. Here is, in brief outline, the gist of our analysis. \textbf{WA} picks up on an apparent tension between applying \textit{classical} mathematics to a certain domain (e.g. an empirical domain) and simultaneously endorsing a \textit{non-classical} logic in reasoning about that same domain. It then purports to elevate the latter tension to an actual inconsistency, which is supposed to corner the deviant logician into a freezing dilemma: either she is to rebuild mathematics in her own logic, or she is forced to forgo the applicability of mathematics to empirical domains, thereby losing (at least some of) the explanatory power characteristic of the natural and social sciences. As this outline already suggests, the hard work in \textbf{WA} lies in elevating that tension to an inconsistency. But it is precisely at this point that we part ways with Williamson, as we argue precisely that the alleged tension, and thus the alleged inconsistency, does not exist. 

The paper is structured as follows. Section 2 starts by carefully reconstructing \textbf{WA} with the purpose of clarifying what exactly are the aforementioned undesirable consequences threatening deviant logics, as well as Williamson's reasons for believing these threats are significant. Section 3 then focuses on a particular logic targeted by \textbf{WA}, namely on quantum logic, and shows that at least in this case there is no threat at all, due to the falsehood of a crucial premise of the argument. After arguing in Section 4 that the same premise fails also in the case of a logic of vagueness, we generalize our criticism and draw some general lessons concerning the relationship between applied logic and applied mathematics. We defer a closer analysis of this relationship to further work. 

\section{Williamson's argument}

A typical deviant logician's response to the fact that mathematical practice seems to involve a constant appeal to classical-logical principles is well expressed in the following passage, quoted by Williamson as well: ``[M]athematical practice is consistent with these reasoning steps [i.e. the ones present in mathematical reasoning] being instances of \textit{mathematical} principles of reasoning, not generalizable to all other discourses. \textit{A fortiori}, they may very well be principles of reasoning that are permissible for mathematics, but not for theorizing about truth.'' (Hjortland 2017, 652–3) This response suggests that just because certain reasoning principles are validated in mathematical discourse, the deviant logician may contend that the same principles do not need to extend to other regions of discourse, e.g. those concerned with truth or, for that matter, mountains or quantum-mechanical phenomena. While this is tenable as long as the mathematical and the non-mathematical are kept mutually isolated, Williamson believes that the applicability of mathematics to non-mathematical domains, as exemplified in the natural and social sciences, raises a problem. Let us now briefly lay out this problem, as Williamson understands it. 

Let $\mathcal{D}$ be a set of terms denoting the non-mathematical objects that the deviant logician takes to require a non-classical treatment (e.g. heaps, electrons, or metalinguistic terms). Also, suppose the deviant logician wants to keep mathematics classical. Williamson argues that this is problematic for several reasons. First of all, the deviant logician cannot consistently substitute elements of $\mathcal{D}$ for variables in the statements of some classical mathematical theorems. For example, the many-valued logician cannot apply the theorem $\forall x,y (x=y \vee x \neq y)$ to, say, mountains without contradicting her thesis that the identity of mountains is possibly undetermined. Furthermore, the same reason prevents the deviant logician from appealing to an isomorphism holding between the objects denoted by $\mathcal{D}$, on the one hand, and some classical purely mathematical objects, on the other hand: e.g. an isomorphism between a collection of mountains and a pure set (as defined by standard ZFC set theory). Lastly, not only does the deviant logician encounter troubles with the application of classical mathematical \textit{theorems}, but she also faces problems with the use of mathematical \textit{reasoning} in science, for she cannot substitute elements of $\mathcal{D}$ for variables used in classical derivations from scientific hypotheses. Since Williamson takes such limitations to imply the impossibility of applying classical mathematics to the non-mathematical objects denoted by the terms in $\mathcal{D}$, and since he also takes it to be uncontroversial that classical mathematics can be applied - and is routinely applied - to the world outside pure mathematics, he concludes that ``in a non-classical world, pure mathematics is no safe haven for classical logic.'' (2018, 21)

As a consequence, the deviant logician would be forced to rebuild mathematics within her own logic. For example, a quantum logician cannot make sense of the applicability of classical mathematics to quantum phenomena, unless she recovers that mathematics in a quantum-logical framework. However, the task of rebuilding mathematics within a new logic is notoriously difficult, and Williamson thinks it would rather unavoidably lead to \textit{ad-hoc} premises, just like the ones that appear in Hartry Field's recapture of the least number principle in a logic without the law of excluded middle (Field 2008).\footnote{Actually, Williamson finds Field's recovery strategy unacceptable for a couple of reasons: unlike the least number principle, which is routinely derived from mathematical induction, Field's recovered version of that principle is a postulate; also, that version is underivable from a suitable schema for mathematical induction. In any case, such reasons for blocking classical recapture are inconsequential to our purposes in this paper.} This would, in turn, make the scientific explanations that involve applications of classical mathematics -- and it is hard to find many that do not -- more costly, since they must include extra assumptions, e.g., the \textit{ad-hoc} premises required for classical recapture (2018, 19). Therefore, so Williamson concludes, even though the path of rebuilding mathematics within a non-classical logic is the one that a staunch deviant logician ought to take, it nevertheless leads to an abductively implausible view -- implausible, because the usual path trodden by the classical logician offers a more elegant and less explanatorily costly alternative.

In order to make sure that the structure of the argument is clear, let us summarize it as follows. Again, $\mathcal{D}$ is a set of terms denoting non-mathematical objects that are tentatively taken to require a non-classical logic $\mathcal{L_D}$. Also, suppose mathematics can be (or even has been) successfully applied to the objects referred to by $\mathcal{D}$. Then \textbf{WA} is just the following dilemma:

\smallskip

1. $\mathcal{L_D}$ can be taken either \textit{to extend} to mathematics or \textit{not to extend} to mathematics.


2. If $\mathcal{L_D}$ is taken \textit{not to extend} to mathematics, then it is inconsistent to hold that mathematics can be applied to the objects referred to by $\mathcal{D}$. 




3. If $\mathcal{L_D}$ is taken \textit{to extend} to mathematics, then most explanations in science become more costly.




4. Therefore, maintaining that reasoning about the objects referred to by $\mathcal{D}$ can be adequately captured by $\mathcal{L_D}$ is either inconsistent or abductively implausible.

\smallskip

Is \textbf{WA} sound? We think not.\footnote{Neil Tennant has recently defended a similar position, by arguing however that Premise 3 fails in the case of his preferred substructural logic (Tennant 2022). Let us also note that \textbf{WA} may as well be countered by maintaining, as Quine suggested, that changing the logic changes the meaning of the logical constants. On this understanding of deviant logics and of how they relate to classical logic, it is clear that no tension with classical mathematics can arise in the first place. However, Williamson is not concerned with such a view (2018, 12), nor are we in this paper (but see [reference removed]).} As already mentioned, we will problematize the first horn of the dilemma, i.e., more particularly, we will show that Premise 2 turns out to be false when the argument is restricted to the case of quantum logic.

\section{Quantum logic avoids \textbf{WA}}

One of the logics that satisfy the presuppositions of \textbf{WA}, and that are thus allegedly threatened by it, is quantum logic (QL). Williamson suggests this much: ``[S]ince quantum mechanics applies mathematics ubiquitously, those who propose non-distributive quantum logic as a serious alternative to classical logic are not excused from the need to reconstruct mathematics on the basis of their quantum logic.'' (2018, 20) Obviously, he is here pushing the second horn of \textbf{WA} against the quantum logician, while at the same time noting in all fairness the best attempt so far, by Michael Dunn, to classically recapture mathematics within non-distributive quantum logic (Dunn 1980). Presumably, Williamson assumes that the quantum logician could stand no chance against the first horn of \textbf{WA}. That is, he appears to assume as established already that it is inconsistent to maintain, on the one hand, that reasoning about quantum phenomena can be adequately captured by QL, \textit{and} to believe at the same time, on the other hand, that classical mathematics can be applied to quantum phenomena. But as we will now show, this alleged inconsistency is clearly dissolved once a closer look is taken at both of these hands.

\subsection{How mathematics is applied to quantum phenomena}\label{math appl}

It is a matter of scientific practice that, in their modelling of quantum phenomena, physicists regularly apply classical mathematical tools from linear algebra, functional analysis, group theory, probability theory, and other standard mathematical theories. Before sketching what some of these applications look like, let us say something about the internal structure of the involved models.\footnote{In this section, for convenience, we closely follow Richard Healey's terminology and formulation of quantum-mechanical models (Healey 2017). Needless to say, by doing so we are not committing ourselves to his, or to any other, particular philosophical interpretation of quantum mechanics. For a critical discussion of Healey's quantum inferentialism, see [reference removed].} A non-relativistic quantum-mechanical model, or simply a \textit{quantum model}, can be denoted by an ordered list $\Theta\equiv \langle \mathcal{H}, T, \mathcal{A}, \psi, \mu_{\phi} \rangle$, where the listed symbols stand for the following mathematical objects:
\begin{itemize}
    \item $\mathcal{H}$ denotes a potentially infinite-dimensional complex Hilbert space, commonly taken to represent the state space of the modelled system (analogously to the phase space that is used for the same purpose in classical mechanics);
    \item $T \subseteq \mathbb{R}$ is an interval on the real numbers, intended to denote a time interval; 
    \item $\mathcal{A}$ is a set of self-adjoint operators on $\mathcal{H}$, corresponding to the dynamical variables associated to the modelled system, e.g. if the system under consideration is an electron, then $\mathcal{A}$ will include operators corresponding to its position, momentum, spin, etc. According to standard quantum mechanics, the values that can be taken by a dynamical variable are contained in the spectrum Spec($A$) of its corresponding operator $A$;
    \item $\psi: T \rightarrow \mathcal{H}$ is a function that assigns a ``quantum state'' $\ket{\psi_t} \in \mathcal{H}$ to each time instant denoted by $t$. In models of isolated systems, i.e., systems that are taken not to interact with any external environment, $\psi$ satisfies Schrödinger's equation $i \partial_t \ket{\psi_t}=H\ket{\psi_t}$, for some Hamiltonian operator $H \in \mathcal{A}$;
    \item for each unit-norm vector $\ket{\phi} \in \mathcal{H}$, $\mu_{\phi}$ is a measure on the set of subspaces of $\mathcal{H}$. In particular, for an arbitrary subspace $S \subseteq \mathcal{H}$: $\mu_{\phi}(S)=\bra{\phi}P_S\ket{\phi}$, where $P_S$ is the projector on $S$.
\end{itemize}

Thus, a quantum model is an ordered collection of classical mathematical objects. But how, more exactly, are these objects applied in the modelling of concrete experiments? 

When a physicist is confronted with the task of constructing a quantum model of a certain experiment of duration $T$, she will start by assigning a Hilbert space $\mathcal{H}$, a set of self-adjoint operators $\mathcal{A}$ and an ``initial'' quantum state $\ket{\psi_{t_0}} \in \mathcal{H}$ to the system involved in the experiment. These assignments are of course not arbitrary, but highly constrained by the conditions set up in the given experiment. For example, if at the beginning of the experiment, a device is activated that is capable of emitting a single electron of a certain energy, this warrants an assignment of particular objects $\mathcal{H}, \mathcal{A}$ and $\ket{\psi_{t_0}}$ to the electron. Furthermore, in order to characterize the dynamical evolution of the target system, the modeller assigns a particular Hamiltonian operator $H \in \mathcal{A}$ that generates the function $\psi$ according to the Schrödinger equation. Up to this point, the physicist has been merely assigning mathematical objects to her experiment, according to the rules and constraints of standard quantum theory. The important step that brings the model closer to empirical reality is provided by the so-called \textit{Born rule}, briefly explained as follows. 

Suppose that at a certain ``final'' time $t_f$, the physicist performs a measurement of a collection of dynamical variables associated to the system of interest (e.g. the position and the spin of an electron). Let $\mathbf{A}\equiv \left\{A_1,...,A_n \right\} \subset \mathcal{A}$, for some $n \in \mathbb{N}$, be the set of self-adjoint operators that correspond to these variables. Importantly, according to quantum theory, there is no possible experiment that could implement a simultaneous measurement of so-called ``incompatible'' variables, i.e. variables associated to mutually non-commuting operators (e.g., a particle's position and momentum). Consequently, the operators in $\mathbf{A}$ obey $[A_i,A_j]=0$, for all $i,j=1...,n$. Next, for each $i$, let $\mathcal{M}^{A_i}_{\Delta^{(i)}} \subseteq \mathcal{H}$ be the eigenspace corresponding to a subset $\Delta^{(i)} \subseteq \text{Spec}(A_i)$ of $A_i$'s spectrum. The Born rule then prescribes (or predicts) the probability $\text{Pr}^{\mathbf{A}}_{t_f}(\Delta^{(1)},...,\Delta^{(n)})$ for the outcomes of the measurement of the $n$ dynamical variables performed at time $t_f$ to be contained respectively in subsets $\Delta^{(1)},...,\Delta^{(n)}$:
\begin{equation}
    \text{Pr}^{\mathbf{A}}_{t_f}(\Delta^{(1)},...,\Delta^{(n)})=\mu_{\psi_{t_f}}(\mathcal{M}^{A_1}_{\Delta^{(1)}}...\mathcal{M}^{A_n}_{\Delta^{(n)}})=\bra{\psi_{t_f}}P^{A_1}_{\Delta^{(1)}}...P^{A_n}_{\Delta^{(n)}}\ket{\psi_{t_f}},
\end{equation}
where $P^{A_i}_{\Delta^{(i)}}$ is the projector on subspace $\mathcal{M}^{A_i}_{\Delta^{(i)}}$, for each $i$. The Born rule thus enables the physicist to use the abstract quantum model -- more precisely, its Hilbert space measures $\mu_{\phi}$ -- to generate (classical) probability distributions $\text{Pr}^{\mathbf{A}}_{t_f}$ that can then be compared to statistical data models extracted from actual experiments. 

This brief review of the application of classical mathematics to standard quantum mechanics will be enough to drive our push back against the first horn of \textbf{WA}. Before we can do so, let us run through the way QL can be used to capture reasoning about quantum phenomena.

\subsection{How QL captures reasoning about quantum phenomena}\label{QL appl}

Despite the empirical success of quantum mechanics, i.e., the fact that quantum-mechanical modelling of the sort just described yields unprecedented predictive and explanatory power with respect to the microscopic world, various problems, such as the measurement problem, Bell's nonlocality, contextuality, and others, leave the interpretational conundra of quantum mechanics still far from resolved. Some realist interpretations urge us, for instance, that the lesson we ought to take from the empirical success of quantum mechanics is that there are invisible particles that occasionally interact non-locally, or that our ``world'' is just one among the many that partake in the Everettian ``multiverse'' (Lewis 2016). Such metaphysical lessons have to a considerable extent overshadowed a series of attempts -- most starkly exemplified in (Putnam 1968) -- at arguing that what quantum mechanics is primarily teaching us is that the ``one true logic'', or at least the logic we ought to adopt in our reasoning about the microscopic world, is Birkhoff and von Neumann's non-distributive quantum logic (or some of its variants).\footnote{See Birkhoff and von Neumann 1936. For more recent presentations, see e.g. Dalla Chiara \textit{et al.} 2004. Some physicists continue to assert that ``real quantum mechanics is not so much about particles and waves as it is about the nonclassical logical principles that govern their behavior.'' (Susskind and Friedman 2014, 236)} With its violation of the distributive law motivated by the peculiarities of quantum mechanics, QL looks like the perfect target of \textbf{WA}. On behalf of the quantum logician, let us recall the details of the quantum-mechanical considerations that led to the adoption of a deviant, non-distributive logic. 

First, we should get clearer on what we mean by `QL' in this paper. Admittedly, there are many formulations and presentations of QL, and albeit most of them agree in their core rejection of the distributive law, some of them yield different relations of logical consequence, which makes it more accurate to speak of QLs, in the plural (Rédei 1998). Nevertheless, since the subtleties surrounding these differences do not affect the point of this paper, we will for simplicity focus on one QL and refer to it in the singular -- thereby meaning that we focus on one particular relation of logical consequence. We will, however, construct this relation by following two different routes, one ``abstract'', the other ``operationalist'', thus called for reasons that will become clearer in a moment. Even though it would be sufficient for our purposes to stick to only one of these routes -- for, recall, our aim is to deny the inconsistency of adopting QL in the quantum domain while still maintaining classical mathematics as applicable therein -- we choose to present both routes, because they will lead to different ways of refuting Premise 2 in \textbf{WA}, which will prove insightful for the more general discussion at the end of the paper.

Consider once again an experiment involving a quantum system, say a particle, and its corresponding quantum model $\Theta$. Since in what follows the dynamical aspects of the model, i.e. those aspects that have to do with time dependence, will be irrelevant, let us turn our attention to its ``kinematical'' part, $\tilde{\Theta}\equiv\langle \mathcal{H}, \mathcal{A}, \mu_{\phi} \rangle$. As stated in the previous section, each element $A$ of $\mathcal{A}$ corresponds to a dynamical variable (e.g., position or momentum) associated to the system under consideration, and each of these variables can take values in its operator's spectrum Spec($A$). Now, roughly speaking, the quantum logician urges us that there is something peculiar about the logical relations that obtain between certain sentences that refer to the dynamical variables associated to our system. The canonical set of sentences that the quantum logician asks us to focus on is the set of ``experimental sentences'' $\mathcal{E}$, i.e., the set that can be generated by conjoining, disjoining and negating so-called ``elementary sentences'', each of which specifies the range of values taken by a variable associated to our quantum system. More precisely, an elementary sentence is of the general form ``$\text{S}^{A}_{\Delta}$'', which is shorthand for ``The value of the variable corresponding to operator $A$ lies in interval $\Delta$'', with $A \in \mathcal{A}$ and $\Delta \subseteq \text{Spec}(A)$. When the informal, natural language semantics just given to the elements of $\mathcal{E}$ is regimented in terms of a rigorous, formal semantics, the road bifurcates into the aforementioned abstract and operationalist routes.\footnote{Cf. Rédei 1998, 68-74 and Malament 2002 for the abstract route; Putnam 1968, 192-197 and Bacciagaluppi 2009, 54-55 for the operationalist one.}

Let us take the abstract route first. This involves interpreting the sentences of $\mathcal{E}$ in terms of an abstract structure borrowed from the mathematical machinery employed in quantum models. Namely, let $h$ be a function (an ``interpretation'') that maps each sentence in $\mathcal{E}$ to a closed subspace of the Hilbert space $\mathcal{H}$ pertaining to our kinematical model $\tilde{\Theta}$. Thus, each elementary sentence of the form ``$\text{S}^{A}_{\Delta}$'' is mapped to a closed subspace $\mathcal{M}^{A}_{\Delta}$, such that the following rules for the QL-connectives determine the mappings of all composite sentences:
\begin{equation}
    \begin{split}
        &h(a \land b)=h(a)\cap h(b)\\
        &h(a \vee b)=h(a) \sqcup h(b)\\
        &h(\lnot a)=h(a)^{\perp},
    \end{split}
\end{equation}
where $a$ and $b$ are arbitrary sentences in $\mathcal{E}$. Here the operations `$\cap$' and `$\sqcup$' denote intersection and linear span, respectively, whereas the `$^{\perp}$'-operation outputs the subspace orthogonal to its input. Now we are ready to introduce the quantum-logical consequence relation as follows. For any $\Gamma \cup \left\{a\right\} \subset \mathcal{E}$, the premises $\Gamma$ are said to quantum-logically imply conclusion $a$, or $\Gamma \vDash_{QL} a$, if and only if 

\begin{equation}\label{log cons ab}
    \bigcap_{\gamma \in \Gamma} h(\gamma) \subseteq h(a).
\end{equation}

It can be easily checked that `$\vDash_{QL}$' disobeys the classical-logical consequence relation, in that there exist triples of sentences $a,b,c \in \mathcal{E}$ for which distributivity does not hold, i.e., they are such that $a\land(b \vee c) \centernot \vDash_{QL} (a \land b) \vee (a \land c)$.

The abstract route, just sketched, owes its name to its defining the semantics of QL in purely mathematical terms, without any further extra-mathematical interpretation added on top of it. Let us now take the alternative, operationalist route. This involves an alternative semantics, which is explicitly defined in terms of idealized operational procedures that may in principle be carried out in a laboratory. First, let $\mathcal{T}$ be a possibly uncountable set, whose elements are taken to denote \textit{idealized tests}, i.e. idealized operational procedures that may hypothetically be applied to physical systems. We assume that the application of any test to any physical system may result in only one of the two mutually exclusive outcomes: either the system passes the test or it does not. Next, let a function $\tau: \mathcal{E} \rightarrow \mathcal{T}$ associate one idealized test to each experimental sentence. In particular, each elementary sentence of the form ``$\text{S}^{A}_{\Delta}$'' is mapped to the test that consists in an idealized measurement of the dynamical variable associated to $A$, such that if the test is applied, it is passed by a system if and only if the outcome of the measurement is contained in $\Delta$. Furthermore, tests associated to compound sentences are determined via the following algorithm, with $a,b \in \mathcal{E}$ henceforth being arbitrary sentences:
\begin{itemize}
    \item $\tau(\lnot a)$ is passed by a system with certainty (i.e. unit probability) if and only if the same system would certainly fail to pass $\tau(a)$;
    \item $\tau(a \land b)$ is passed by a system with certainty if and only if the same system would certainly pass both $\tau(a)$ and $\tau(b)$;
    \item $\tau(a \vee b)$ is passed by a system with certainty if only if the same system would certainly pass either $\tau(a)$ or $\tau(b)$.
\end{itemize}

Now, let us define the relation of logical consequence on the sentences in $\mathcal{E}$ with the use of their associated tests: heuristically, we will say that the set of premises, $\Gamma$, QL-implies conclusion $b$, if it is the case that any quantum system that passes each test $\tau(\gamma)$ with certainty, for all $\gamma \in \Gamma$, also passes $\tau(b)$ with certainty. In order to formalize this, let us introduce a set of valuations $\mathcal{V}\equiv \left\{ V_{\phi}| \forall \phi \in \mathcal{H}, \text{s.t.} \braket{\phi|\phi}=1\right\}$, where $\mathcal{H}$ is the Hilbert space assigned to the quantum system under consideration. Each $V_{\phi}: \mathcal{T}\rightarrow \left\{t,f\right\}$ is a bivalent truth-valuation that assigns one of the two truth values to each test, as follows: $V_{\phi}(g)=t$ if and only if a quantum system prepared in quantum state $\ket{\phi}$ would pass test $g$ with certainty, where $g \in \mathcal{T}$ is arbitrary. Whether a system passes a certain test with certainty is in turn determined by the outcome of hypothetical applications of quantum models to idealized tests. This means that the value of $V_{\phi}(g)$ is determined by the probability distribution generated from an appropriate quantum model of the test $g$ performed on a system in state $\ket{\phi}$. Without getting into more details about the relation between tests, valuations and quantum models, once the definition of valuations is in place, the logical consequence relation follows canonically. For any $\Gamma \cup \left\{b\right\} \subset \mathcal{E}$, $\Gamma \vDash_{QL} b$ if and only if:
\begin{equation}\label{log cons op}
    \forall V \in \mathcal{V}: \quad \text{if} \quad \left(\forall \gamma \in \Gamma,  V(\gamma)=t \right) \quad \text{then} \quad V(b)=t.
\end{equation}
As it can be easily verified, the hereby constructed logical consequence relation coincides with the one found above at the end of the abstract route (i.e., the one given in Eq. \eqref{log cons ab}), which is the reason we denote both of them by the same symbol `$\vDash_{QL}$'. Now that we have caught a glimpse, cursory though this may be, into quantum mechanics and into QL, we are ready to get back to \textbf{WA}, and in particular ready to challenge its second premise.

\subsection{Why there is no inconsistency}\label{no inconsistency}

When restricted to the case of QL, Premise 2 of \textbf{WA} can be paraphrased as: ``It is inconsistent to simultaneously hold that (i) QL adequately captures reasoning about quantum phenomena, and that (ii) classical mathematics can be applied to quantum phenomena.'' As we will now show, this premise is false.

Recall that our quantum logician holds that the logical relations between experimental sentences require a non-distributive logic. As noted above, experimental sentences are obtained by logically compounding elementary sentences, such as $\text{S}^{(X)}_{[1,2]}$, which reads ``The value of an electron's position lies in the interval $[1,2]$.'' But what does this have to do with the possibility of applying classical mathematics to the quantum domain? Do the quantum logicians' claims concerning experimental sentences have any empirical bearing that could prevent the application of classical mathematics in quantum modelling? There is no unqualified reply to these questions, because the informal, natural language semantics does not pinpoint the precise content of a sentence like $\text{S}^{(X)}_{[1,2]}$. Does this sentence specify the electron's location, just like the statement ``We are currently in Vienna'' specifies our location? Or is it rather just shorthand for the conditional statement ``If a measurement were performed, its output would indicate the electron's position somewhere in the interval $[1,2]$''? Needless to say, these two possible readings do not have the same empirical significance: the former, i.e., the statement ``The value of the electron's position lies in the interval $[1,2]$.'' may be understood as a metaphysical claim about properties instantiated by invisible objects in the world, without any implications on what we may observe in a performed experiment, whereas the second reading, as a conditional statement, has a direct bearing on possible observations in experiments that may be carried out in a laboratory. 

Now, notice that the first formal semantics, which led us via an abstract route to the quantum-logical consequence relation, involves only uninterpreted classical mathematics, and is thus entirely compatible with any of the two readings of experimental sentences. In other words, the quantum logicians' experimental sentences, together with a formal semantics inspired by the mathematical formalism of quantum mechanics, may as well have no empirical bearing at all, thus avoiding any conflict with the applicability of classical mathematics to quantum phenomena. So this abstract route already provides a counterexample to Premise 2, but it is admittedly rather cheap, in that it effectively isolates experimental sentences from (the talk about) empirical reality. For example, it makes no connection between $\text{S}^{(X)}_{[1,2]}$ and possible outcomes of experiments performed on electrons. In contrast, the second formal semantics, which led us via an operationalist route to the same quantum-logical consequence relation, makes such a connection, as it equates the meaning of experimental sentences with operations that can in principle be performed in a laboratory. This route thus exposes the quantum logician to a possible inconsistency with the applicability of classical mathematics. Indeed, now that her experimental sentences are taken to have an empirical bearing, the non-distributive logical relations between them may appear to be in tension with the classicality of mathematics employed in quantum modelling. Nevertheless, as we will see presently, this tension is merely apparent.

The quantum logician that walks along the operationalist route maintains that certain sentences that concern the outcomes of operational procedures performable in a laboratory obey a non-distributive logic. So in order to see whether this conflicts with the applicability of classical mathematics to the quantum domain, we need to determine what this very applicability implies with respect to the logical relations between experimental sentences. First, consider a generic class of quantum models $\Theta_{\alpha}=\langle \mathcal{H}, \mathcal{A}, T, \psi, \mu_{\phi} \rangle$, where the index $\alpha$ indicates that the ``initial'' quantum state of its corresponding model is $\ket{\alpha} \in \mathcal{H}$, i.e. that $\ket{\psi_{t_0}}=\ket{\alpha}$. With the help of the Born rule, each of these models can be used to describe a variety of experiments, as mentioned in Section \ref{math appl}. Secondly, consider a particular experiment $T^{\mathbf{A}}_{\alpha}$, which consists of the measurement of a set of dynamical variables associated to operators $\mathbf{A}\equiv \left\{A_1,...,A_n\right\}\subset \mathcal{A}$ on a quantum system prepared in state $\ket{\alpha}$. Recall that according to quantum theory, any set of operators corresponding to simultaneously measurable variables is necessarily commutative, which implies that $[A_i,A_j]=0$, for all $i,j=1,...,n$. Lastly, let $\text{Pr}^{\mathbf{A}}_{\alpha}$ be the probability distribution generated from the quantum model $\Theta_{\alpha}$ in an application to $T^{\mathbf{A}}_{\alpha}$. Now, we can show that the classicality of $\text{Pr}^{\mathbf{A}}_{\alpha}$ implies that the logical relations among a certain subset of experimental sentences obey classical logic. 

Let $S^{(i)}_{\Delta}$ be the elementary sentence ``The value of the variable associated to $A_i$ lies in $\Delta$'', where $\Delta$ is an arbitrary subset of $\text{Spec}(A_i)$. Furthermore, let $\mathcal{E}_{\mathbf{A}}\subset \mathcal{E}$ be the subset of experimental sentences generated by logically compounding the elementary sentences $S^{(i)}_{\Delta}$. Now, notice that the aforementioned experiment $T^{\mathbf{A}}_{\alpha}$ can be used to furnish idealized tests $\tau(S^{(i)}_{\Delta})$ for the elementary sentences $S^{(i)}_{\Delta}$. Indeed, saying that a quantum system prepared in state $\ket{\alpha} \in \mathcal{H}$ passes test $\tau(S^{(i)}_{\Delta})$ with certainty is equivalent to asserting that
\begin{equation}
   \text{Pr}^{\mathbf{A}}_{\alpha}(G^{(i)}_{\Delta})=1,
\end{equation}
where $G^{(i)}_{\Delta}$ is shorthand for $(\Omega_1,...\Omega_{i-1},\Delta,\Omega_{i+1},...,\Omega_n)$, with $\Omega_j\equiv$ Spec($A_j$). Furthermore, the quantum logician's operational semantics for disjunction and conjunction implies the following relations between the tests associated to these connectives and the probability distribution $\text{Pr}^{\mathbf{A}}_{\alpha}$. Namely, a quantum system prepared in state $\ket{\alpha}$ 
\begin{itemize}
    \item passes test $\tau(S^{(i)}_{\Delta_i} \land S^{(j)}_{\Delta_j})$ if and only if $\text{Pr}^{\mathbf{A}}_{\alpha}(G^{(i)}_{\Delta_i}\cap G^{(j)}_{\Delta_j})=1$,
    \item passes test $\tau(S^{(i)}_{\Delta_i} \vee S^{(j)}_{\Delta_j})$ if and only if $\text{Pr}^{\mathbf{A}}_{\alpha}(G^{(i)}_{\Delta_i}\cup G^{(j)}_{\Delta_j})=1$,
\end{itemize}
for arbitrary $i,j=1,...,n$, and for $\Delta_i \subseteq \text{Spec}(A_i)$ and $\Delta_j \subseteq \text{Spec}(A_j)$. Note that, being a classical probability distribution, $\text{Pr}^{\mathbf{A}}_{\alpha}$ obeys the distributive identity
\begin{equation}
   \text{Pr}^{\mathbf{A}}_{\alpha}(G^{(i)}_{\Delta_i}\cap(G^{(j)}_{\Delta_j}\cup G^{(k)}_{\Delta_k}))=\text{Pr}^{\mathbf{A}}_{\alpha}( (G^{(i)}_{\Delta_i}\cap G^{(j)}_{\Delta_j})\cup (G^{(i)}_{\Delta_i} \cap G^{(k)}_{\Delta_k})), 
\end{equation}
which in turn, together with the operational definition of the QL-consequence relation \eqref{log cons op}, implies the validity of the following argument
\begin{equation}
    S^{(i)}_{\Delta_i}\land(S^{(j)}_{\Delta_j}\vee S^{(k)}_{\Delta_k}) \vDash_{QL} (S^{(i)}_{\Delta_i}\land S^{(j)}_{\Delta_j})\vee (S^{(i)}_{\Delta_i} \land S^{(k)}_{\Delta_k}).
\end{equation}

Therefore, by just carefully following her own operationalist route, the quantum logician seems to be forced to assert the validity of the classical law of distributivity. This is a consequence of the classicality of the probability distributions used in modelling quantum-mechanical phenomena. 

However, let there be no misunderstanding: the quantum logician is forced to assert the validity of the classical law of distributivity \textit{only} for the subset of sentences $\mathcal{E}_{\mathbf{A}}$ generated by logically compounding the elementary sentences  $S^{(i)}_{\Delta}$, which refer exclusively to dynamical variables associated to elements in the set of commutative operators $\mathbf{A}\subset \mathcal{A}$. Does this present a problem for the quantum logician? 

Definitely not! It is perfectly consistent to maintain that the set $\mathcal{E}$ of all experimental sentences requires a non-classical logic, while admitting that there are subsets like $\mathcal{E}_{\mathbf{A}}$, which nonetheless validate classical logic. Importantly, it is no accident that we started our construction from a set of commutative operators: had we considered a non-commutative set, such as $\left\{X,P\right\}\subset \mathcal{A}$, where $X$ and $P$ are operators corresponding to, say, a particle's position and momentum, the construction would have been blocked from the get-go. This is because there is no (idealized) experiment, adequately modelled by a joint classical probability distribution, that could simultaneously measure variables whose corresponding operators are mutually non-commuting. Briefly put, since any application of a quantum model can only be used to generate probability distributions over values of quantum-mechanically compatible variables, it follows that the classicality of such distributions can force only some subsets of experimental sentences $\mathcal{E}_{\mathbf{A}} \subset \mathcal{E}$, for any set of commutative operators $\mathbf{A}$, to obey classical logic. This is the case indeed \textit{according to QL}. Therefore, the alleged tension between QL and the applicability of classical mathematics in quantum mechanics does not exist.

But the fact that this tension does not exist is not surprising at all.\footnote{That QL does not conflict with the classicality of quantum-mechanical models has already been suggested e.g. by Dickson (2001, S283), as well as by Putnam (1968), Bacciagaluppi (2009), and others. Nevertheless, we found it worthwhile, for the sake of our counterargument, to make the lack of this conflict, as well as the reasons that underlie such a lack, more explicit.} To insist that it does, as \textbf{WA} has suggested, is actually a curiosity of sorts. Having explained why neither the abstract, nor the operationalist quantum logician is threatened by \textbf{WA}, we should recall that the operationalist's quantum-logical consequence relation is itself constructed on the basis of quantum theory and its mathematical formalism. The logical relations between experimental sentences, as explained in Section \ref{QL appl}, are related to properties of tests performable in principle on quantum systems. But these tests and their properties are not freely posited by a speculative logician: they are instead explicitly defined in terms of quantum models! This is why we find it rather odd that an argument such as \textbf{WA} has been levelled against QL, for the latter has been explicitly motivated by, and formally constructed on the basis of, the mathematical formalism of quantum theory. That \textbf{WA} is unsound in the case of QL is, thus, not an accident. 

QL was but one target. As noted at the outset, there are others. Since the motivation and construction of many alternative logics are entirely independent of mathematical models employed in science, one might feel inclined to think that these fall prey more readily to \textbf{WA}. Many-valued logics, for example, have been thought to adequately capture reasoning about vague objects and phenomena, and they can certainly be constructed independently of the scientific models of these objects and phenomena. Nonetheless, our analysis will show that, when \textbf{WA} is restricted to this case, Premise 2 fails once more.

\section{Deviant logics of vagueness also avoid WA}
Vagueness and its related paradoxes constitute a paramount motivator for non-classical logics. Indeed, it is often maintained that reasoning about heaps, mountains and baldness requires the rejection of some classical-logically valid principle or rule, such as the law of excluded middle or the Cut rule. Nevertheless, those who endorse deviant logics of vagueness contend that the latter's non-classicality does not infect the realm of mathematics: ``Presumably excluded middle holds throughout ordinary mathematics, and indeed whenever vagueness or indeterminacy is not at issue.'' (Field 2008, 101). In this case, however, \textbf{WA} allows the contention only if the language of mathematics could be kept isolated from the non-mathematical vocabulary infected by vagueness. But then it further points out that the mathematical and the vague are actually not two independent realms, as exemplified in (natural and social) scientific applications of mathematics to non-mathematical domains, which are hardly ever immune to vagueness and borderline cases.\footnote{That empirical domains are infected with vagueness can be illustrated by questions like: What is the maximum distance between a proton and an electron such that they still constitute a hydrogen atom? If a series of tumultuous political events count as a revolution, at what precise moment in time, counted in milliseconds, did it begin?} \textbf{WA} thus concludes that the deviant logician of vagueness ought to reconstruct mathematics in her non-classical logic, if she wishes not to lose the explanatory power of scientific-mathematical models. 

Williamson acknowledges that the deviant logician may attempt to flee the dilemma ``on the grounds that when applying mathematics in science one can pretend that the relevant non-mathematical terms are precise, because the gap between this pretence and reality is irrelevant for scientific purposes.'' (2018, 20) However, this acknowledgement is immediately followed by criticism, according to which ``[i]f there is vagueness in the relevant terms, and the true mathematics for vague terms is non-classical (because the correct logic for them is), then one would expect ``vagueness errors'' in the predictions.'' (\textit{loc. cit.}) Williamson is, thus, claiming that the ``gap between pretence and reality'' is quantifiable and can lead to systematic errors due to the misalignment between the classical mathematics used in a scientific model and the non-classical mathematics purportedly inherent to the modelled target objects or phenomena (i.e. their purported ``true mathematics''). In what follows, we will present a counterexample to Williamson's claims, by considering a scientific application of classical mathematics to an object apparently infected with vagueness, and will explicitly show that reasoning about the latter in accordance with a deviant logic is perfectly consistent with the use of classical logic in the mathematical model of that object. More specifically, we examine a classical mathematical model, based on Newtonian mechanics, of a planet's movement.

Consider a team of scientists and engineers, whose aim is to land a space-probe on a certain planet $X$. Suppose that their strategy for executing the task is guided by the use of a classical mathematical model constructed on the basis of Newton's laws, which is intended to represent the motion of the planet and its surroundings. Without getting into unnecessary details of what Newtonian models may generally look like, suppose that our team's model consists simply of a ``mass-density distribution'' function $\rho: \mathbb{R}^3\times \mathbb{R}^+ \rightarrow \mathbb{R}^+$, which satisfies a set of differential equations developed by our team (roughly, by means of a rather sophisticated application of Newton's second law). The interpretation of the model is the following. For each location denoted by a vector $\vec{x} \in \mathbb{R}^3$ and each time instant denoted by $t \in \mathbb{R}^+$, the value $\rho(\vec{x},t) \in \mathbb{R}^+$ denotes the density of mass present at the respective location and time instant. The total quantity of mass present in a certain volume $V \subset \mathbb{R}^3$ at a time instant $t$ is, thus, given by the integral $\int_V \rho(\vec{x},t) d^3x$. Besides representing the matter that constitutes planet $X$ and its immediate surroundings, the distribution function $\rho$ may also describe the matter pertaining to other bodies - e.g. other planets, stars or dust - that the team takes to be relevant for their task. Suppose that our team also specifies precisely which pieces of matter pertain to planet $X$ and which do not, thereby modelling the planet as if it had clearly defined spatial boundaries. Mathematically, this assumption states that for each time instant $t$, there is a subset $V^X_t \subset \mathbb{R}^3$, such that the function $\rho(\vec{x}, t)$, restricted to the points $\vec{x} \in V^X_t$ represents the mass density of the planet at time instant $t$. 

However, as already mentioned above, the term `planet' seems to be paradigmatically vague, in that it appears that it either fails to refer to a unique material object, or that its referent lacks sharp boundaries. This purported (semantic or ontological) vagueness spreads to statements in which the term `planet' figures, such as $S(X,y)\equiv$ ``Material object $y$ is part of planet $X$'': if $y$ is a rock trapped a few kilometres under $X$'s surface, then $S(X,y)$ is undeniably true, but if $y$ consists in cosmic dust that has just landed on $X$'s surface, then $S(X,y)$ seems to be neither true nor false, thus qualifying as the deviant logician's potential counterexample to the classical law of excluded middle or to the principle of bivalence. The deviant logician, on the one hand, seemingly portrays the mathematical activities that our scientists and engineers partake in as a game of pretense, in which the team is representing planet $X$ \textit{as if} the latter had clear boundaries, when it in fact does not. Williamson, on the other hand, denies that our team is playing any games, and holds that $X$ clearly has sharp boundaries, since the lack of sharpness would otherwise manifest itself empirically: our team would then certainly notice the vagueness of the term `planet', or they at least should do so, if their execution of the task is to be as successful as possible. Let us now show why we think Williamson's take is incorrect, and more generally, why the deviant logician does not run into any inconsistencies in cases like the one just described.

As sketched above, our team's model consists of a mathematical function $\rho$ that is taken to represent the matter that constitutes the planet, its surroundings, and potentially other celestial bodies that may be of relevance. Now, this model may be employed, for example, to determine whether some matter $y$ will be present in some volume $V'$ at time $t'$, which may indeed prove to be useful information if our team is to successfully coordinate the landing of its space-probe (our team might want to minimize contact with additional matter - e.g. with dust - as the latter could ruin the space-probe's equipment). Importantly, notice that it is entirely irrelevant for our team's purposes whether $y$ is part of the material body commonly referred to as `planet $X$', `cosmic dust $Y$', `star $Z$', etc., or whether there is any such fact of the matter at all. Rather, what is important for our probe landing is how much matter is roughly present within the space-time region that is of interest. More generally, the empirical adequacy of model $\rho$ is entirely independent of whether some of its elements are associated to a seemingly vague term, such as `planet', `dust', or `probe', and of whether any such association is to be considered as absolutely (in)correct -- as Williamson would have it -- or as a pragmatically justified output of a game of pretension - as we may put it on behalf of the deviant logician. In other words, empirical adequacy is independent of whether statements such as $S(X,y)$ are true, false, both or neither.

Here is another way of making this point. To see that any views concerning the truth-value (or lack thereof) of statements such as $S(X,y)$ are inconsequential for the applicability of our model $\rho$ is to recognize the conceptual possibility of adopting an instrumentalist stance, which conceives of the mathematical model as a mere device for generating predictions about data extracted from observations or experiments performed on the modelled celestial bodies. According to this stance, the classicality of the mathematical model can, if anything, only force the extracted data to be classical, but has no say on what logic is adequate for reasoning, in natural language, about the sources that produced the data. In particular, the classicality of the mathematical model cannot impose classical-logical relations onto  statements like ``This cloud of cosmic dust is part of planet \textit{X}''. Even though such a view of scientific models may be explanatorily unsatisfactory, it is definitely not an inconsistent one, as \textbf{WA} would have it. Premise 2 thus fails once again: there is no inconsistency between (i) holding that reasoning about planets, their material constituents, their color and size, as they appear in our ``manifest image'', is adequately captured by a deviant logic, and (ii) maintaining that classical mathematics can be used in modelling the dynamics of the matter that constitutes planets and their surroundings. Let us now conclude by briefly indicating what we take to be a deeper problem with \textbf{WA}, which we think might explain the two symptomatic failures of Premise 2 so far discussed.

\section{Conclusion: A deeper problem with WA?}

We have argued that, whether the motivation and construction of a deviant logic is dependent on mathematical models actually employed in science (as in the case of non-distributive quantum logic) or not (as in the case of non-classical logics of vagueness), Premise 2 in \textbf{WA} fails. This would appear to suggest that there is some deeper problem with Williamson's attack on deviant logics, which comes to surface in both cases. But what could that be? 

Although classical logic (in the context of one's mathematical model) and the deviant logics (in the context of one's communications outside of the mathematical model) are both deployed in reasoning about one and the same system, closer analysis showed that their targets may effectively differ. More specifically, we have seen that, in quantum mechanics, a classical mathematical model accounts for quantities that arise in a particular experimental context, whereas quantum-logical reasoning concerns relations between experimental propositions, which refer to quantities that arise in potentially different experimental contexts. Similarly, the classical mathematical model used by our team in landing their space-probe accounts for quantifiable aspects of the material constituents of the modelled planet and of its surroundings, whereas ordinary natural-language discourse about that same planet disregards such aspects. 

This suggests the following general diagnosis. Let \textbf{T} be one's target system or phenomenon, and \textbf{M} be a classical-logical mathematical model applicable to \textbf{T}. The diagnosis is that there are many cases in which, even though \textbf{M} may be used to account for certain aspects of \textbf{T}, there may be other aspects of \textbf{T} that are articulable via some discourse \textbf{D} whose structure is adequately captured by a non-classical logic. In other words, the gulf between \textbf{M} and \textbf{T} is too wide for the mere fact that \textbf{M} is applied successfully to \textbf{T} to significantly restrict the logical structure of any meaningful assertoric discourse \textbf{D} about that same object of investigation \textbf{T}. Such a restriction may incidentally arise \textit{only} with further assumptions about what the success of a mathematical model implies about the modelled object or phenomenon. 

For instance, an ontic structural realist, as the one portrayed by Ladyman \textit{et al.} (2007), may maintain that at least in some cases the successful application of a mathematical model to \textbf{T} implies that \textbf{T} itself is ``already structured'' in some particular way, which may in turn restrict the logical structure of other kinds of discourse about \textbf{T}. Conversely, other accounts of the use of mathematics in science, such as the one articulated by Bas van Fraassen (2010), may sever such restrictions. Indeed, while on van Fraassen's view, a successful mathematical model is taken to be isomorphic (or rather homomorphic) to the data model generated from a target system (or more precisely, to the surface model generated from a data model), the relation between the data model and the target system is in turn determined by ``pragmatic factors'' -- there need be no kind of morphism between them. This is another way of saying that ``assertions of isomorphism are context-sensitive'' (\textit{op. cit.}, 382), where the different contexts that such assertions are sensitive to allow for different data models to be constructed from one and the same target system. If this is correct, then the logic of the data model need not be the same as the logic used in reasoning about other aspects of the target system. So it is not really that surprising, if van Fraasen's view is adopted, that one can apply \textbf{M} to $\textbf{T}$, while holding onto the conviction that reasoning about other aspects of $\textbf{T}$ is adequately captured with a non-classical logic. Or, with reference to our own examples, it is not that surprising that one can apply classical-logical quantum-mechanical models to quantum systems, while holding onto the conviction that drawing inferences from experimental sentences requires a non-classical logic, just as one can apply classical-logical Newtonian models to planetary movements, while drawing non-classical inferences from statements about qualitative aspects of such movements. 

The upshot is that, whether one takes the ontic structuralist realist's, van Fraassen's, or some other position, any attempt of characterizing the relationship between applied logic and applied mathematics, such as the one embodied in $\mathbf{WA}$, ought to make explicit its background theoretical account of applications of mathematics in science. Without moving this background into the foreground, there can be no significant tension arising between applied logic and applied mathematics, let alone an inconsistency.


\section*{References}

Bacciagaluppi, G. (2009). Is Logic Empirical? In Gabbay D., D. Lehmann, and K. Engesser (eds.) \textit{Handbook of Quantum Logic}, Elsevier, Amsterdam, 49-78. 

\smallskip

Birkhoff, G. and J. von Neumann (1936). The Logic of Quantum Mechanics. In \textit{Annals of Mathematics}, 37, 823-843.

\smallskip

Dalla Chiara, M., R. Giuntini, and R. Greechie (2004) \textit{Reasoning in Quantum Theory. Sharp and Unsharp Quantum Logics}, Kluwer, Dordrecht.

\smallskip





Dunn, J. M. (1980). Quantum Mathematics. In \textit{Proceedings of the the Philosophy of Science Association}, 512--531.

\smallskip

Field, H. (2008). \textit{Saving truth from paradox},  Oxford University Press.

\smallskip



Healey, R. (2017). \textit{The quantum revolution in philosophy}. Oxford University Press.

\smallskip

Hjortland, O. T. (2017). Anti-exceptionalism about logic. \textit{Phil. Studies}, 174, 631-658.

\smallskip

Hlobil, U. (2021). Limits of abductivism about logic. \textit{Philosophy and Phenomenological Research, 103}(2), 320-340.

\smallskip

Martin, B., \& Hjortland, O. T. (2022). Anti-exceptionalism about logic as tradition rejection. \textit{Synthese, 200}(2), 148.

\smallskip

Ladyman, J., Ross, D., \& Spurrett, D. (2007). \textit{Every thing must go: Metaphysics naturalized}. Oxford University Press.
\smallskip

Lewis, P. J. (2016). \textit{Quantum ontology: A guide to the metaphysics of quantum mechanics}. Oxford University Press.
\smallskip

Malament, D. B. (2002). Notes on Quantum Logic. Unpublished lecture notes.

\smallskip

Priest, G. (2016). Logical disputes and the a priori. \textit{Logique et Analyse}, (236), 347-366.

\smallskip

Putnam, H. (1968). Is Logic Empirical? In R. S. Cohen and M. W. Wartofsky (eds.) \textit{Boston Studies in the Philosophy of Science}, 5, Dordrecht, Reidel, 216-241, reprinted as ``The logic of quantum mechanics'', in \textit{Mathematics, Matter and Method. Philosophical Papers}, 1, Cambridge University Press, 174-197.

\smallskip






\smallskip

Rédei, M. (1998). \textit{Quantum Logic in Algebraic Approach}, Dordrecht, Kluwer Academic Publishers.

\smallskip

Susskind, L. and A. Friedman (2014). \textit{Quantum Mechanics. The Theoretical Minimum}, New York, Basic Books.

\smallskip

Tennant, N. (2022). On the Adequacy of a Substructural Logic for Mathematics and Science, \textit{The Philosophical Quarterly}, 72, 4, 1002-1018. 

\smallskip



van Fraassen, B. C. (2010) \textit{Scientific Representation. Paradoxes of Perspective}, Oxford University Press.

\smallskip

Williamson, T. (2017). Semantic paradoxes and abductive methodology. \textit{Reflections on the Liar}, 325-346.

\smallskip

Williamson, T. (2018). Alternative logics and applied mathematics. \textit{Philosophical Issues, a supplement to Noûs}, 28(1), 399-424.

\end{document}